\documentstyle[12pt,epsf]{article}
\newcommand{\n}{\hspace*{-2.5mm}}
\newcommand{\lsim}{\;\rlap{\lower 3.5 pt \hbox{$\mathchar \sim$}} \raise 1pt
 \hbox {$<$}\;}

\begin{document}

\title{\vskip-3cm{\baselineskip14pt
\centerline{\normalsize\hfill MPI/PhT/96--65}
\centerline{\normalsize\hfill hep-ph/9610456}
\centerline{\normalsize\hfill July 1996}
}
\vskip1.5cm
Virtual Top-Quark Effects on the $H\to b\bar b$ Decay at Next-to-Leading Order
in QCD}
\author{{\sc K.G. Chetyrkin}\thanks{Permanent address:
Institute for Nuclear Research, Russian Academy of Sciences,
60th October Anniversary Prospect 7a, Moscow 117312, Russia.},
{\sc B.A. Kniehl, and M. Steinhauser}\\
{\normalsize Max-Planck-Institut f\"ur Physik (Werner-Heisenberg-Institut),}\\
{\normalsize F\"ohringer Ring 6, 80805 Munich, Germany}\\ \\
}
\date{}
\maketitle
\begin{abstract}
By means of a heavy-top-quark effective Lagrangian, we calculate the
three-loop corrections of ${\cal O}(\alpha_s^2G_FM_t^2)$ to the $H\to b\bar b$
partial decay width of the standard-model Higgs boson with intermediate mass
$M_H\ll2M_t$.
We take advantage of a soft-Higgs theorem to construct the relevant
coefficient functions.
We present our result both in the $\overline{\mbox{MS}}$ and on-shell schemes
of mass renormalization.
The $\overline{\mbox{MS}}$ formulation turns out to be favourable with regard
to the convergence behaviour.
We also test a recent idea concerning the na\"\i ve non-abelianization of QCD.

\medskip
\noindent
PACS numbers: 12.38.-t, 12.38.Bx, 14.65.Ha, 14.80.Bn
\end{abstract}
\newpage

One of the central questions of elementary particle physics is whether nature
makes use of the Higgs mechanism of spontaneous symmetry breaking to endow the
particles with their masses.
In the framework of the minimal standard model, the Higgs boson, $H$, is the
missing link to be experimentally discovered in order to complete our
understanding of mass generation.
So far, all attempts to detect the Higgs boson on its mass shell have been in
vain, with the effect that the mass range $M_H\le65.2$~GeV has been ruled out
at the 95\% confidence level (CL) \cite{gri}.
However, experimental precision tests of the standard electroweak theory are 
sensitive to the Higgs boson via quantum corrections.
A recent global fit \cite{alc} has yielded $M_H=149{+148\atop-82}$GeV together 
with a 95\% CL upper bound of 550~GeV.

A Higgs boson with $M_H\lsim135$~GeV decays dominantly to $b\bar b$ pairs.
This decay mode will be crucial for Higgs-boson searches at LEP~2, the
Fermilab Tevatron after the installation of the Main Injector, a
next-generation $e^+e^-$ linear collider, and a future $\mu^+\mu^-$ collider.

The present knowledge of quantum corrections to the $H\to b\bar b$ partial
decay width has recently been reviewed in Ref.~\cite{mod}.
At one loop, the electroweak \cite{nuc} and QCD \cite{bra} corrections are
known for arbitrary masses.
In the limit $M_H\ll2M_t$, in which we are interested here, the terms of
${\cal O}(X_t)$, where $X_t=\left(G_FM_t^2/8\pi^2\sqrt2\right)$, tend to be
dominant.
They arise in part from the renormalizations of the Higgs wave function and
vacuum expectation value, which are independent of the quark flavour \cite{cha}.
In the case of bottom, there is an additional non-universal ${\cal O}(X_t)$
contribution \cite{nuc}, which partly cancels the flavour-independent one.
At two loops, the universal \cite{hll} and bottom-specific \cite{hbb}
${\cal O}(\alpha_sX_t)$ terms are available.
Furthermore, the first \cite{gor} and second \cite{sur} terms of the
expansion in $M_b^2/M_H^2$ of the ${\cal O}(\alpha_s^2)$ five-flavour QCD
correction have been found.
As for the top-quark-induced correction in ${\cal O}(\alpha_s^2)$, the full
$M_t$ dependence of the non-singlet (double-bubble) contribution \cite{kni} as
well as the first four terms of the $M_H^2/M_t^2$ expansion of the singlet
(double-triangle) contribution \cite{lar} have been computed.
At three loops, the ${\cal O}(\alpha_s^3)$ non-singlet correction is known in
the massless approximation \cite{che}.
Furthermore, the $H\to gg$ decay width receives a ${\cal O}(\alpha_s)$
correction due to the $b\bar bg$ final state which may also be interpreted as
a ${\cal O}(\alpha_s^3M_H^2/M_b^2)$ correction to the $H\to b\bar b$ decay
width \cite{zer}.

We have taken the next step by evaluating the three-loop
${\cal O}(\alpha_s^2X_t)$ corrections to the $H\to b\bar b$ decay width.
In this letter, we report the key results.
A detailed description of our analysis together with a discussion of the
${\cal O}(\alpha_s^2X_t)$ corrections to the less important $H\to q\bar q$ 
decay widths, where $q=u,d,s,c$, will be presented elsewhere \cite{cks}.

We now outline our procedure.
We construct a $n_f=5$ effective Yukawa Lagrangian, ${\cal L}_Y^{\rm eff}$, by
integrating out the top quark.
This Lagrangian is a linear combination of dimension-four operators acting in
QCD with $n_f=5$ quark flavours, while all $M_t$ dependence is contained in
the coefficient functions.
We then renormalize this Lagrangian and, by exploiting the
renormalization-group (RG) invariance of the energy-momentum tensor, rearrange
it in such a way that the renormalized operators and coefficient functions are
separately independent of the renormalization scale, $\mu$.
The final result for ${\cal L}_Y^{\rm eff}$ exhibits the following structure:
\begin{equation}
\label{eff}
{\cal L}_Y^{\rm eff}=-2^{1/4}G_F^{1/2}H\left(1+\bar\delta_{\rm u}\right)
\left[{\cal C}_1\left[O_1^\prime\right]
+\sum_q\left({\cal C}_{2q}\left[O_{2q}^\prime\right]
+{\cal C}_{3q}\left[O_{3q}^\prime\right]\right)\right],
\end{equation}
where $q$ is runs over $u,d,s,c,b$ and the primed objects refer to QCD with
$n_f=5$.
Here, $\bar\delta_{\rm u}$ is the universal correction resulting from the
renormalizations of the Higgs wave function and vacuum expectation value.
The square brackets denote the renormalized counterparts of the bare $n_f=5$
operators
\begin{equation}
O_1^\prime=\left(G_{a\mu\nu}^{0\prime}\right)^2,\qquad
O_{2q}^\prime=m_q^{0\prime}\bar q^{0\prime}q^{0\prime},\qquad
O_{3q}^\prime=\bar q^{0\prime}\left[\frac{i}{2}
\left(\stackrel{\rightarrow}{D}\hspace{-.85em}/{}^{0\prime}
-\stackrel{\leftarrow}{D}\hspace{-.85em}/{}^{0\prime}\right)
-m_q^{0\prime}\right]q^{0\prime},
\end{equation}
where $G_{a\mu\nu}$ is the colour field strength, $D_\mu$ is the covariant 
derivative, and the superscript 0 labels bare fields and parameters.
${\cal C}_1$, ${\cal C}_{2q}$, and ${\cal C}_{3q}$ are the respective
renormalized coefficient functions.
Notice that $\left[O_{3q}^\prime\right]$ vanishes by the fermionic equation of
motion, so that ${\cal C}_{3q}$ will not appear in our final result.
Nevertheless, the inclusion of $O_{3q}^\prime$ is indispensable for the
determination of ${\cal C}_1$ and ${\cal C}_{2q}$.
The RG-improved formulation thus obtained provides a natural separation of the
$n_f=5$ QCD corrections at scale $\mu=M_H$ and the top-quark-induced $n_f=6$
corrections at scale $\mu=M_t$, in the sense that the final result for the
$H\to b\bar b$ decay width will not contain logarithms of the type
$\ln(M_t^2/M_H^2)$ if the $n_f=5$ and $n_f=6$ corrections are expanded in
$\alpha_s^{(5)}(M_H)$ and $\alpha_s^{(6)}(M_t)$, respectively.

In contrast to the two-loop ${\cal O}(\alpha_sX_t)$ case \cite{hbb}, where it
was sufficient to consider just one term in ${\cal L}_Y^{\rm eff}$, we need to
take into account three types of operators and to allow for them to mix under
renormalization.
The mixing terms are related to the ${\cal O}(\alpha_s^2)$ double-triangle
contribution considered in Ref.~\cite{lar} and extend the latter to
${\cal O}(\alpha_s^2X_t)$.

\begin{figure}[ht]
   \leavevmode
 \begin{center}
 \begin{tabular}{cc}
   \epsfxsize=3.0cm
   \epsffile[189 361 423 565]{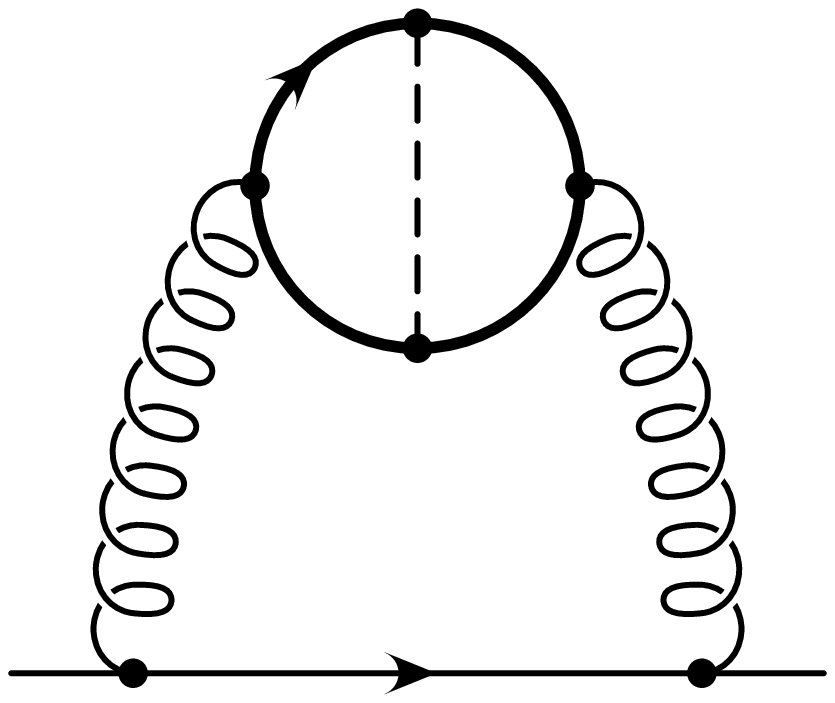}
&
   \epsfxsize=3.0cm
   \epsffile[189 361 423 565]{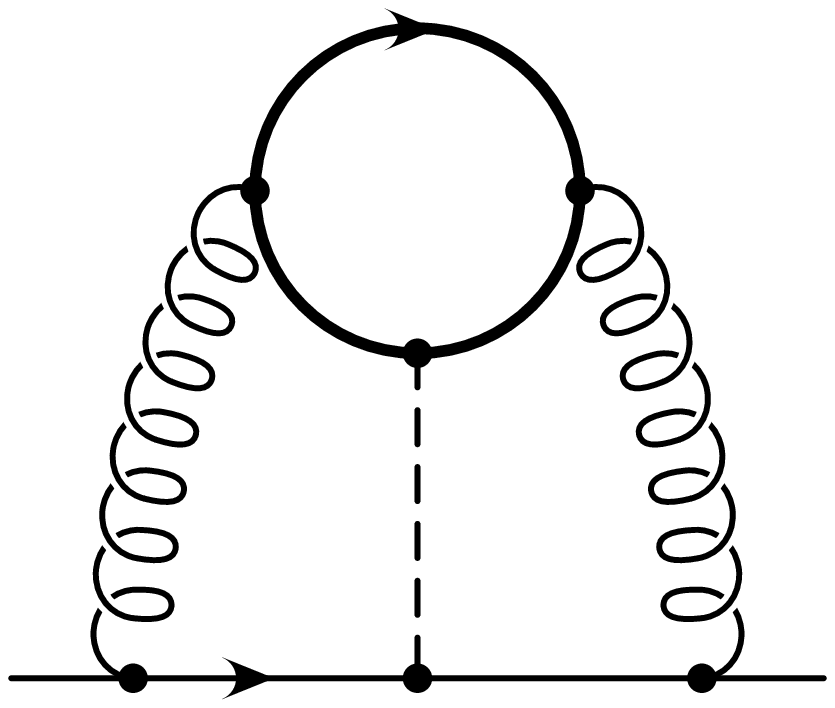}
 \end{tabular}

 \begin{tabular}{ccc}
   \epsfxsize=4.0cm
   \epsffile[165 267 447 419]{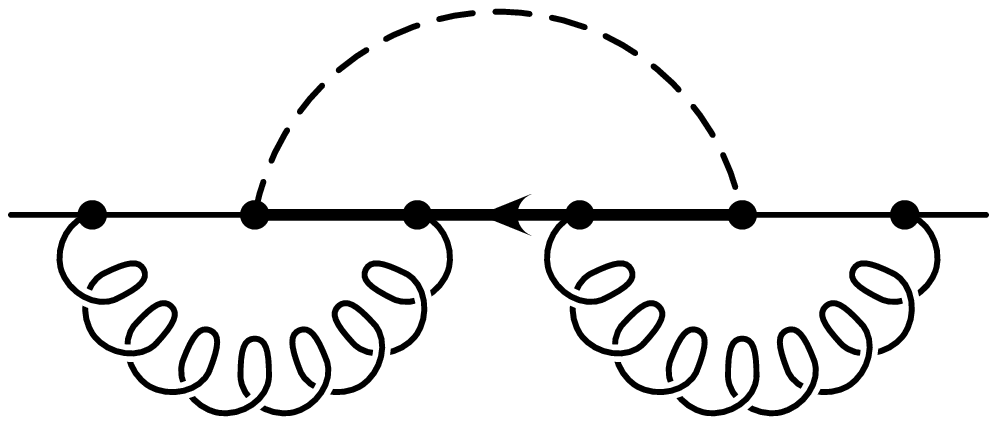}
&
   \epsfxsize=4.0cm
   \epsffile[165 267 447 419]{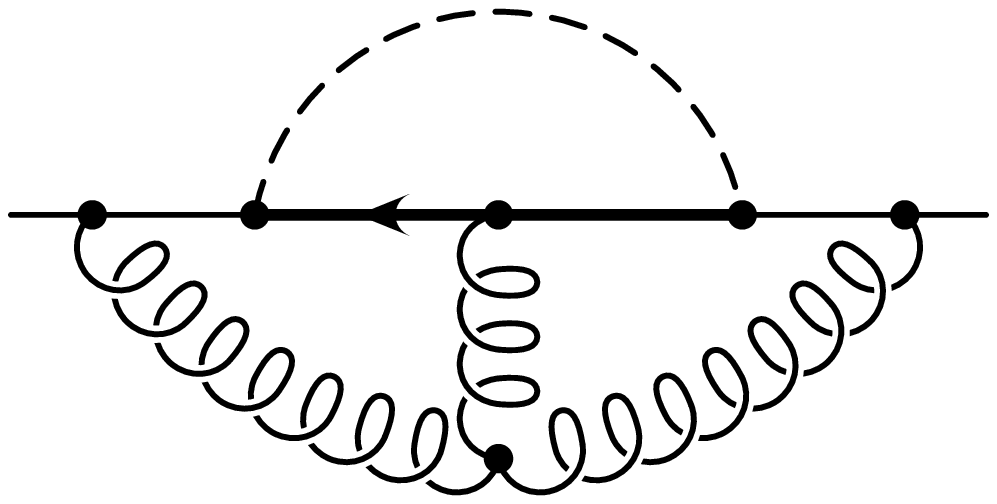}
&
   \epsfxsize=4.0cm
   \epsffile[165 267 447 419]{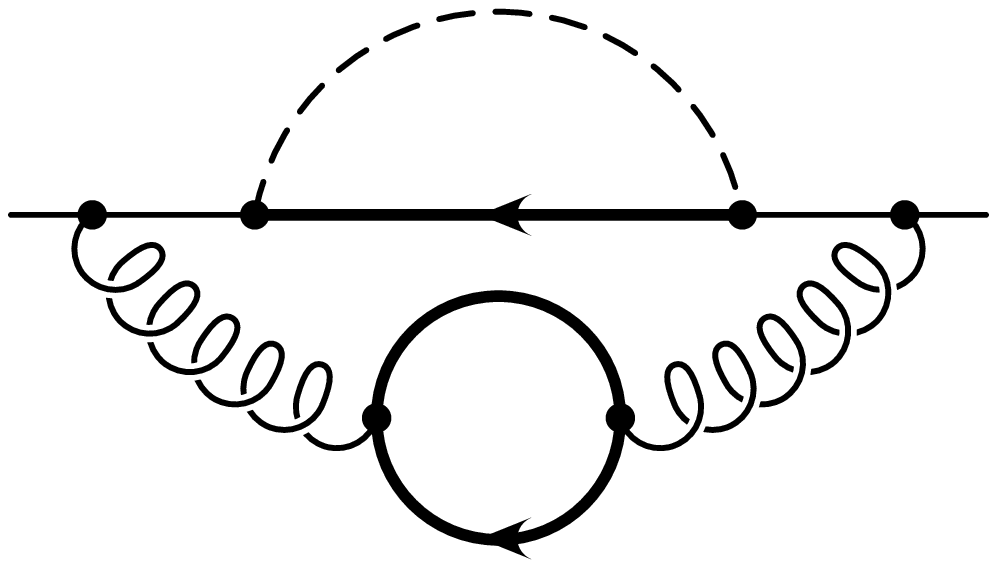}
 \end{tabular}
  \caption{\label{dia} Typical diagrams generating universal and
non-universal ${\cal O}(\alpha_s^2X_t)$ corrections to
$\Gamma\left(H\to b\bar b\right)$.
Bold-faced (dashed) lines represent top quarks (Higgs or Goldstone bosons).}
 \end{center}
\end{figure}

Similarly to Ref.~\cite{hbb}, we can take advantage of the higher-order
formulation \cite{spi} of a well-known soft-Higgs theorem \cite{let} to
simplify the calculation of the coefficient functions.
This allows us to relate a huge number of three-loop three-point diagrams to
a manageable number of three-loop two-point diagrams.
Specifically, we have to compute 24 irreducible three-loop two-point diagrams
for $q\ne b$ and, in addition, 54 ones for $q=b$.
Typical examples are depicted in Fig.~\ref{dia}.
Such a theorem is not available for the gauge interactions, which might
explain why three-loop ${\cal O}(\alpha_s^2X_t)$ corrections have not yet been
calculated for the $Z\to q\bar q$ decay widths, including the important case of
$Z\to b\bar b$, which has recently attracted much attention in connection with
the so-called $R_b$ anomaly.

From Eq.~(\ref{eff}) we may derive an expression for the $H\to b\bar b$ decay
width, appropriate for $M_H\ll2M_t$, which accommodates all presently known
corrections, including the new ones of ${\cal O}(\alpha_s^2X_t)$.
It reads
\begin{equation}
\label{hbb}
\Gamma\left(H\to b\bar b\right)=\Gamma_{b\bar b}^{\rm Born}
\left[\left(1+\Delta_b^{\rm QED}\right)
\left(1+\left.\Delta_b^{\rm weak}\right|_{X_t=0}\right)
\left(1+\Delta_b^{\rm QCD}\right)\left(1+\Delta_b^t\right)
+\Xi_b^{\rm QCD}\Xi_b^t\right].
\end{equation}
Here,
\begin{equation}
\label{born}
\Gamma_{b\bar b}^{\rm Born}=\frac{3G_FM_Hm_b^2}{4\pi\sqrt2}
\left(1-\frac{4m_b^2}{M_H^2}\right)^{3/2}
\end{equation}
is the Born result including the full mass dependence.
As is well known \cite{bra}, we may avoid the appearance of large logarithms
of the type $\ln(M_H^2/m_b^2)$ in the QCD correction, $\Delta_b^{\rm QCD}$, by
taking $m_b$ in Eq.~(\ref{born}) to be the $\overline{\mbox{MS}}$ mass
evaluated with $n_f=5$ quark flavours at scale $\mu=M_H$, $m_b^{(5)}(M_H)$.
Consequently, we may put $m_b=0$ in $\Delta_b^{\rm QCD}$.
We may proceed similarly with the QED correction, $\Delta_b^{\rm QED}$, which
then takes the form
\begin{equation}
\Delta_b^{\rm QED}=\frac{17}{36}\,\frac{\alpha(M_H)}{\pi},
\end{equation}
where $\alpha(\mu)$ is the $\overline{\mbox{MS}}$ fine-structure constant.
In turn, $m_b^{(5)}(M_H)$ is then also shifted by a QED correction \cite{hem}
from the pole mass, $M_b$.
$\left.\Delta_b^{\rm weak}\right|_{X_t=0}$ denotes the weak correction with
the leading ${\cal O}(X_t)$ term stripped off.
If we put $m_b=0$ and consider the limit $M_H\ll2M_W$,
$\left.\Delta_b^{\rm weak}\right|_{X_t=0}$ simplifies to \cite{nuc}
\begin{equation}
\left.\Delta_b^{\rm weak}\right|_{X_t=0}=\frac{G_FM_Z^2}{8\pi^2\sqrt2}
\left(\frac{1}{6}-\frac{7}{3}c_w^2-\frac{16}{3}c_w^4
+3\frac{c_w^2}{s_w^2}\ln c_w^2\right),
\end{equation}
where $c_w^2=1-s_w^2=M_W^2/M_Z^2$ and $M_Z$ ($M_W$) is the $Z$-boson 
($W$-boson) mass.
$\Delta_b^{\rm QCD}$ is the well-known QCD correction for $n_f=5$ \cite{gor},
\begin{equation}
\Delta_b^{\rm QCD}=\frac{17}{3}a_5
+a_5^2\left(\frac{8851}{144}-\frac{47}{6}\zeta(2)-\frac{97}{6}\zeta(3)\right),
\end{equation}
where $a_5=\alpha_s^{(5)}(M_H)/\pi$ and $\zeta$ is Riemann's zeta function,
with values $\zeta(2)=\pi^2/6$ and $\zeta(3)\approx1.202$.
This correction originates from the class of diagrams where the Higgs boson
directly couples to the final-state $b\bar b$ pair.
As is well known \cite{lar}, starting at ${\cal O}(\alpha_s^2)$,
$\Gamma\left(H\to b\bar b\right)$ also receives leading contributions from the
$b\bar b$ and $b\bar bg$ cuts of the double-triangle diagrams where the top
quark circulates in one of the triangles.
In the language of Eq.~(\ref{eff}), where the top quark only appears in the
coefficient functions, this class of contributions is generated by the
interference diagram of the operators $\left[O_1^\prime\right]$ and
$\left[O_{2b}^\prime\right]$.
The absorptive part of this diagram also includes a contribution from the $gg$
cut, which is well known and must be subtracted.
This leads to
\begin{equation}
\Xi_b^{\rm QCD}=a_5\left(-\frac{76}{3}+8\zeta(2)
-\frac{4}{3}\ln^2\frac{m_b^2}{M_H^2}\right).
\end{equation}
The would-be mass singularity proportional to $\ln^2(m_b^2/M_H^2)$ cancels if
the $b\bar b(g)$ and $gg$ decay channels are combined \cite{lar}.

Equation~(\ref{hbb}) is arranged in such a way that the leading $M_t$
dependence is carried by 
\begin{equation}
\label{top}
\Delta_b^t=\left(1+\bar\delta_{\rm u}\right)^2
\left({\cal C}_{2b}\right)^2-1,\qquad
\Xi_b^t=\left(1+\bar\delta_{\rm u}\right)^2{\cal C}_1{\cal C}_{2b},
\end{equation}
where $\bar\delta_{\rm u}$, ${\cal C}_1$, and ${\cal C}_{2b}$ are defined in 
the context of Eq.~(\ref{eff}).
Since we use $m_b^{(5)}(M_H)$ in the $n_f=5$ segments of Eq.~(\ref{hbb}),
it appears natural to also renormalize the top-quark mass in the $n_f=6$ terms
according to the $\overline{\mbox{MS}}$ prescription.
The scale-independent definition $\mu_t=m_t^{(6)}(\mu_t)$ is singled out,
since it eliminates the RG logarithms of the type $\ln(\mu^2/m_t^2)$.
By analogy to $X_t$, we define $x_t=\left(G_F\mu_t^2/8\pi^2\sqrt2\right)$.
We shall return to the $M_t$ formulation later on.
From Ref.~\cite{ks} we know that
\begin{equation}
\bar\delta_{\rm u}=x_t\left[\frac{7}{2}
+a_6\left(\frac{19}{3}-2\zeta(2)\right)-9.598\,a_6^2\right],
\end{equation}
where $a_6=\alpha_s^{(6)}(\mu_t)/\pi$.
An analytic expression for $\bar\delta_{\rm u}$, valid for $\mu$ and $N_c$
arbitrary, may be found in Ref.~\cite{ks}.
As mentioned above, ${\cal C}_1$ and ${\cal C}_{2b}$ are RG invariant per
construction.
Owing to the separation of the $n_f=5$ physics at scale $\mu=M_H$ and the
$n_f=6$ physics at scale $\mu=\mu_t$, they are given by \cite{cks}
\begin{equation}
\label{rgc}
{\cal C}_1=\frac{\alpha_s^{(5)}(\mu_t)\beta^{(5)}(M_H)}
{\alpha_s^{(5)}(M_H)\beta^{(5)}(\mu_t)}C_1,\qquad
{\cal C}_{2b}=\frac{4\alpha_s^{(5)}(\mu_t)}{\pi\beta^{(5)}(\mu_t)}
\left(\gamma_m^{(5)}(\mu_t)-\gamma_m^{(5)}(M_H)\right)C_1+C_{2b},
\end{equation}
where $\beta^{(n_f)}(\mu)=\beta\left[\alpha_s^{(n_f)}(\mu)\right]$ is the
Callan-Symanzik beta function,
$\gamma_m^{(n_f)}(\mu)=$\break
$\gamma_m\left[\alpha_s^{(n_f)}(\mu)\right]$ is the
quark-mass anomalous dimension, and \cite{cks}
\begin{eqnarray}
\label{coe}
C_1&\n=\n&a_6\left(-\frac{1}{12}+\frac{1}{4}x_t\right),
\\
C_{2b}&\n=\n&1+\frac{5}{18}a_6^2
+x_t\left\{-3-7a_6+a_6^2\left[-\frac{4201}{48}+38\zeta(2)-\frac{5}{2}\zeta(3)
+n_f\left(\frac{241}{72}-\frac{4}{3}\zeta(2)\right)\right]\right\}.
\nonumber
\end{eqnarray}
Here, we have displayed the coefficient of $n_f=6$, for reasons which will
become clear later on.
General expressions for $C_1$ and $C_{2b}$, valid for $\mu$ and $N_c$
arbitrary, will be provided in Ref.~\cite{cks}.
If we expand Eq.~(\ref{rgc}) and substitute Eq.~(\ref{coe}), Eq.~(\ref{top}) 
becomes
\begin{eqnarray}
\Delta_b^t&\n=\n&
a_6^2\left(\frac{5}{9}
+\frac{2}{3}\ln\frac{\mu_t^2}{M_H^2}\right)
+x_t\left[1+a_6\left(-\frac{4}{3}-4\zeta(2)\right)
+a_6^2\left(-59.163+\frac{2}{3}\ln\frac{\mu_t^2}{M_H^2}\right)\right],
\nonumber\\
\Xi_b^t&\n=\n&a_6\left(-\frac{1}{12}-\frac{1}{12}x_t\right).
\end{eqnarray}
The appearance of the logarithms $\ln(\mu_t^2/M_H^2)$ witnesses the loss of
the RG improvement inherent in Eq.~(\ref{rgc}).

As a by-product, we may derive from Eq.~(\ref{eff}) a formula for the
$H\to gg$ decay width which includes the ${\cal O}(x_t)$ correction. 
The result is
\begin{equation}
\Gamma(H\to gg)=\frac{2\sqrt2G_FM_H^3}{\pi}
\left(1+\bar\delta_{\rm u}\right)^2({\cal C}_1)^2
=\frac{G_FM_H^3a_6^2}{36\pi\sqrt2}(1+x_t),
\end{equation}
which is in agreement with Ref.~\cite{djo}.

If we express Eq.~(\ref{coe}) in terms of $M_t$, we obtain
\begin{eqnarray}
\label{cos}
C_{2b}^{\rm OS}&\n=\n&1+\frac{5}{18}A_6^2
+X_t\left\{-3+A_6+A_6^2\left[-\frac{65}{3}+52\zeta(2)+4\zeta(2)\ln2
-\frac{7}{2}\zeta(3)
\right.\right.\nonumber\\
&\n\n&{}+\left.\left.
n_f\left(\frac{7}{18}-\frac{10}{3}\zeta(2)\right)\right]\right\},
\end{eqnarray}
where $A_6=\alpha_s^{(6)}(M_t)/\pi$, while $C_1^{\rm OS}$ emerges from $C_1$
by merely replacing $a_6$ and $x_t$ with $A_6$ and $X_t$, respectively.
From Eq.~(\ref{cos}) we read off that the leading ${\cal O}(X_t)$ term
receives the QCD correction factor $\left(1-0.333\,A_6-11.219\,A_6^2\right)$.
We thus recover a pattern similar to the electroweak parameter $\Delta\rho$
\cite{avd} and the corrections $\delta_{\rm u}$, $\delta_{WWH}$, and
$\delta_{ZZH}$ \cite{ks} to the $l^+l^-H$, $W^+W^-H$, and $ZZH$ vertices,
respectively.
In fact, the corresponding QCD expansions in $A_6$ of these four observables
all have negative coefficients which dramatically increase in magnitude as one
passes from two to three loops \cite{ks}.
On the other hand, if the top-quark mass is renormalized in the 
$\overline{\mbox{MS}}$ scheme at scale $\mu=\mu_t$, then the respective QCD
expansions in $a_6$ are found to have coefficients which have variant signs
and nicely group themselves around zero \cite{ks}.
We note in passing that the study of infrared renormalons \cite{gam} offers a
possible theoretical explanation for this observation.
In the case of $C_{2b}$, the QCD correction factor reads
$\left(1+2.333\,a_6+7.032\,a_6^2\right)$.
We conclude that, also in the case of the $b\bar bH$ interaction,
the QCD expansion in the on-shell scheme exhibits a worse convergence 
behaviour than the one in the $\overline{\mbox{MS}}$ scheme.
However, the difference is less striking than in the previous four cases.

Finally, we would like to test Broadhurst's rule concerning the na\"\i ve
non-abelianization of QCD \cite{bro}.
Guided by the observation that the $n_f$-independent term of $\beta_0$ emerges
from the coefficient of $n_f$ by multiplication with $-33/2$, Broadhurst
conjectured that this very relation between the $n_f$-independent term and the
coefficient of $n_f$ approximately holds for any observable at next-to-leading
order in QCD.
In Ref.~\cite{ks}, this rule was applied to $\Delta\rho$, $\delta_{\rm u}$,
$\delta_{WWH}$, and $\delta_{ZZH}$, and it was found that, in all four cases,
the signs and orders of magnitude of the $n_f$-independent terms are correctly
predicted.
Except for $\delta_{ZZH}$, these predictions come, in fact, very close to the
true values.
If we multiply the coefficients of $n_f$ in Eqs.~(\ref{coe}) and (\ref{cos})
with $-33/2$, we obtain $-19.041$ and $84.055$, which has to be compared with
the respective $n_f$-independent terms, $-28.018$ and $64.223$.
Once again, the signs and orders of magnitude of the $n_f$-independent terms,
which are usually much harder to compute than the coefficients of $n_f$, come
out correctly.

\end{document}